\documentclass[twocolumn,floatfix,superscriptaddress,a4paper,showpacs,showkeys,nofootinbib,reprint,prc,noeprint]{revtex4-1}
\usepackage{epsfig}
\usepackage{latexsym}
\usepackage[colorlinks=true,linktocpage=true,linkcolor=blue,citecolor=blue,allcolors=blue]{hyperref}
\usepackage{url}
\usepackage[utf8]{inputenc}
\usepackage{enumerate}
\usepackage{color}
\usepackage{xcolor}

\usepackage{xfrac}

\usepackage{amsmath}
\usepackage{amssymb}

\usepackage[english]{babel}
\usepackage{hyperref}
\usepackage{url}
\usepackage{needspace}
\usepackage{float}
\hypersetup{
   colorlinks=true, 
  linktoc=all,     
  linkcolor=blue,  
}

\begin{document}

 \title{Charmed hadron production from secondary $\overline p + p$ annihilations \\ in p+A reactions at FAIR}

\author{Tom Reichert}
\affiliation{Institut f\"{u}r Theoretische Physik, Goethe Universit\"{a}t Frankfurt, Max-von-Laue-Str. 1, D-60438 Frankfurt am Main, Germany}
\affiliation{Frankfurt Institute for Advanced Studies, Ruth-Moufang-Str. 1,  60438 Frankfurt am Main, Germany}

\author{Jan Steinheimer}
\affiliation{GSI Helmholtzzentrum f\"ur Schwerionenforschung GmbH, Planckstr. 1, D-64291 Darmstadt, Germany}
\affiliation{Frankfurt Institute for Advanced Studies, Ruth-Moufang-Str. 1,  60438 Frankfurt am Main, Germany}

\author{Marcus Bleicher}
\affiliation{Institut f\"{u}r Theoretische Physik, Goethe Universit\"{a}t Frankfurt, Max-von-Laue-Str. 1, D-60438 Frankfurt am Main, Germany}
\affiliation{Helmholtz Research Academy Hesse for FAIR (HFHF), GSI Helmholtzzentrum f\"ur Schwerionenforschung GmbH, Campus Frankfurt, Max-von-Laue-Str. 12, 60438 Frankfurt am Main, Germany}

\date{\today}

\begin{abstract}
We present estimates for the production cross sections of exotic states ($\Lambda_c, \Sigma_c, \Xi_c, D\, \mathrm{and}\, D_s$) from secondary $\overline B + B$ annihilations in p+A reactions from $E_\mathrm{lab}=10-30A$~GeV. We focus specifically on the newly planned  hadron physics program of CBM at FAIR. These estimates for the production of exotic states are based on the achievable number of $\overline B + B$ annihilations and their invariant mass distributions calculated in the UrQMD transport model. 
\end{abstract}

\maketitle

\section{Introduction}
The exploration of exotic hadron states of QCD (Quantum-Chromo-Dynamics) has been among the main goals of hadron physics programs around the world. Most notably are the current results from BES III \cite{Zweber:2009qf,BESIII:2020nbj} in $e^++e^-$ reactions on the exploration of the X,Y,Z states, GlueX \cite{Pentchev:2024ffv,GlueX:2019mkq} with its investigation of charm production near the threshold in $\gamma+p$ reactions and the observation of the $\Xi_{cc}$ at LHCb \cite{LHCb:2017iph}.  The PANDA experiment \cite{PANDA:2009yku,Nerling:2021bxo} at FAIR is supposed to contribute to these investigations using an anti-proton beam with initial momenta of up to 15 GeV. A major aim is to explore the properties of exotic (charmed) hadrons and glue-balls with unprecedented precision.
With the currently expected delay of the PANDA experiment, a hadron physics program running as part of the CBM collaboration is developed as a bridge towards the program with anti-protons. This calls for a careful re-analysis of what part of the $\overline p + p$ program could be achieved in this new set-up. 

The CBM experiment's main focus is on reactions of heavy nuclei to create compressed baryonic matter. Unfortunately, the collision energies for heavy nuclei are limited due to their small $Z/A$ ($Z$ is the proton number, $A$ is the baryon number of the nuclei) to the range $\sqrt{ s_\mathrm{NN}} = 3-5$ GeV, which is below the charm production threshold of $\approx 5$ GeV for the $J / \Psi$ in elementary nucleon-nucleon reactions. 
Multi-step processes in heavy ion reactions that allow for the accumulation of energy in heavy resonances have been suggested as sources for charm production near or even below the threshold of elementary nucleon-nucleon reactions \cite{Steinheimer:2016jjk}. 

However, the collision energy at FAIR can be increased, if one restricts the beam to proton projectiles. Here, beam energies of up to E$_\mathrm{lab}^\mathrm{proton}=30A$~GeV are available, lifting the initial $\sqrt{s_\mathrm{NN}}$ to $\approx 7.7$ GeV. Pioneering studies on charm production in this energy range have been performed e.g. in \cite{Linnyk:2008hp} based on production channels in $N+N$ and $\pi+N$. The cross section for open charm production has been estimated to be in the order of a few $10^{-2} \mu b$ in $N+N$ around $\sqrt{s_\mathrm{NN}} = 7.7$ GeV.

The production of charmed hadrons in $p+p$ reactions is still limited by their threshold energy and it is strongly suppressed even above the threshold \cite{Cassing:2000vx,Linnyk:2008hp,Song:2015sfa}, as the effective energy available for particle production in $NN$ reactions is low, due to the large longitudinal motion of the outgoing baryons. 

In the present work, we want to point to an alternative production mechanism that was intensely explored for PANDA, namely the production of charmed states in $\overline B + B$ annihilations. For the PANDA program, the anti-protons would have been produced in separate $p+A$ reactions with much higher collision rates, then collected and subsequently brought to collision with protons or nuclei. We suggest to use the same target nucleus for both, the production of the anti-baryons and their annihilation. This will allow to include our present knowledge on the production of charmed exotica in $\overline p + p$ annihilations into the $p+A$ program and will further allow to scrutinize the high gluon densities reached in the annihilation process.

\section{The case for $\overline B + B$ annihilations}
As discussed above, the production of new quarks in color fields is usually strongly suppressed. Also in $\overline p + p$ annihilations a similar obervation is expected naively, because according to the Okubo-Zweig-Iizuka (OZI) rule, processes that require quark annihilation followed by the creation of a disconnected quark pair should be highly suppressed. This suppression arises because the process relies on an intermediate state composed purely of gluons, which does not directly couple to the initial valence quarks. Consequently, the expected cross section for these reactions, based on OZI rule arguments, are usually very small  (e.g. in the case $\overline p + p \rightarrow \phi+\phi$ on the order of $nb$). However, experimental results from the JETSET collaboration \cite{JETSET:1998akg} contradict this expectation. JETSET observed an anomalously large cross section of $2–4\, \mu b$ for incoming anti-proton momenta corresponding to center-of-mass energies $\sqrt s \approx 2.12$GeV (slightly above the threshold), which is approximately two orders of magnitude larger than the OZI rule prediction. This substantial violation of the OZI rule suggests the presence of additional dynamical effects beyond simple two-gluon exchange in $\overline p + p$ annihilations. 

Several theoretical mechanisms have been proposed to explain the unexpectedly large cross sections in $\overline p + p$ annihilations:
\begin{itemize}
\item Glueball resonance production \cite{Lipkin:1983kd,Lindenbaum:1984wz}:
The intermediate gluonic state may form a resonant glueball, a bound state of gluons predicted by quantum chromodynamics (QCD). Such a state would naturally couple strongly to gluonic processes, leading to an enhancement in  production. Since gluons are massless spin-1 particles, possible quantum numbers for a two-gluon system include  $0^{++}, 0^{-+}, 2^{++}$. Lattice QCD calculations predict glueball masses in the range of $2.39 - 2.56$ GeV, making them accessible in near-threshold  production experiments.

\item Coupling to broad four-quark states \cite{Ke:2018evd,Lu:2019ira}:
In the case of strangeness, the reaction may proceed via intermediate tetraquark states with significant strangeness content, such as the $\phi(2170)$ and the $X(2239)$. These states can act as intermediate resonances, enhancing  production.

\item Hidden strange or charm quark content in the proton and anti-proton \cite{Ellis:1994ww}:
If the proton and anti-proton wavefunctions contain a significant hidden strange or charm component, the reaction could occur with less suppression than traditionally expected.

\item Baryon/meson exchange in the t- and u-channel \cite{Haidenbauer:2014rva,Haidenbauer:2016pva}: Additional contributions from mesonic and baryonic exchange diagrams in the t- and u-channel could provide alternative reaction pathways, further increasing the cross section.
\end{itemize}

Thus, a key to obtain these exotic states is the creation of a gluon-rich environment. This makes anti-baryon+baryon annihilations a much better environment for the creation of exotic states in comparison to proton+proton reaction. In addition, the full annihilation energy is concentrated in a small volume in contrast to a longitudinally stretched color flux tube created in proton+proton reactions. Another advantage is that quantitative estimates for the production of these novel states in $\overline p + p$ annihilations are available \cite{Haidenbauer:2014rva,Haidenbauer:2016pva}, in contrast to $p+p$ reactions at low energies. 

Therefore, we explore here the possibility to use secondarily produced anti-baryons, which annihilate in the same collision system to create the desired gluon-rich environment. We than fold the energy spectrum of the $\overline B + B$ annihilations with the previously obtained production cross sections of hitherto unexplored hadron channels and predict estimates for the experimental reach within the parameters of the CBM experiment (i.e. E$_\mathrm{lab}^\mathrm{proton}=10-30A$ GeV). One should note that we focus here only on the secondary anti-baryons and do not include the production of exotica from (potentially possible) proton+nucleon interactions, thus, we obtain a lower limit.

\begin{figure} [t!]
    \centering
    \includegraphics[width=\columnwidth]{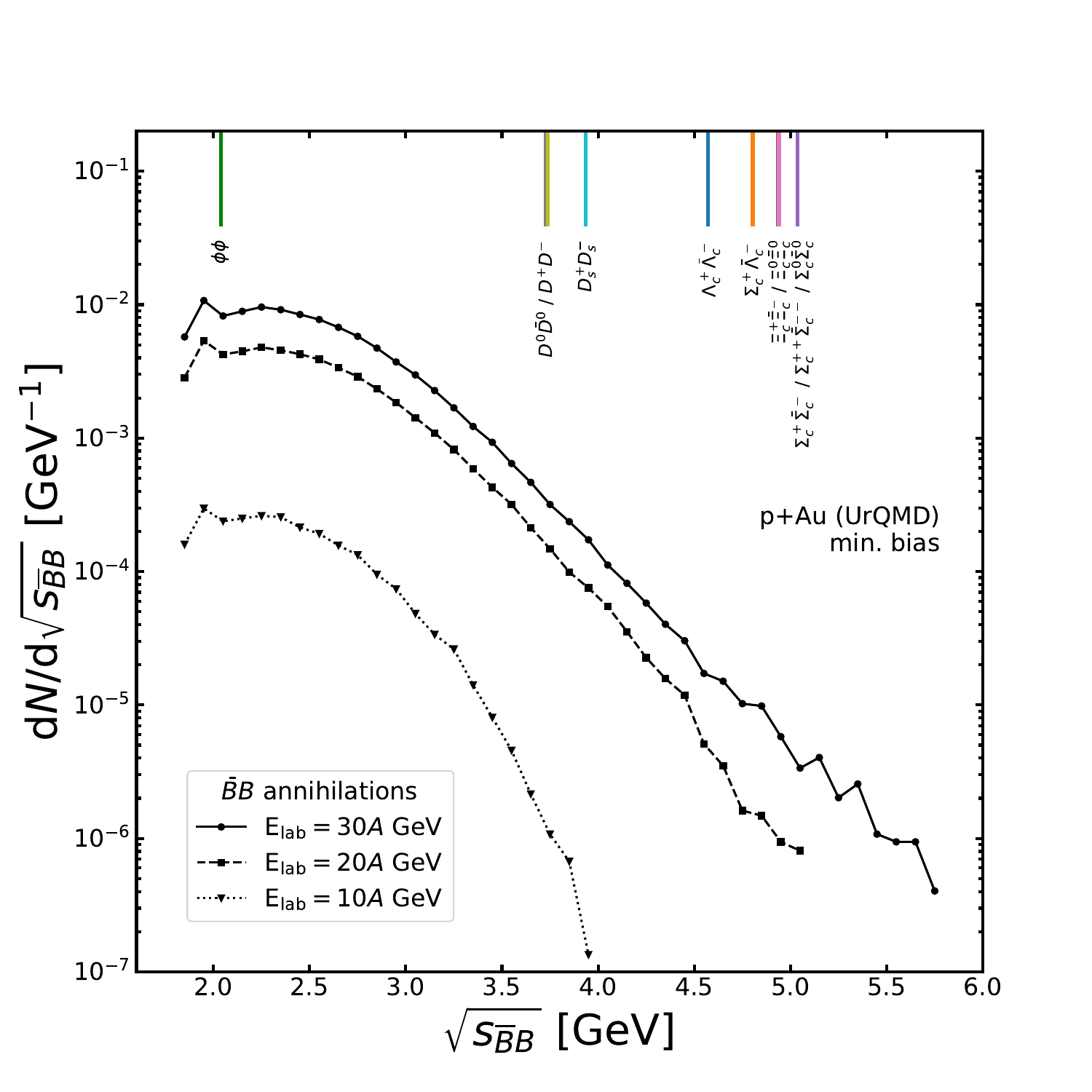}
    \caption{[Color online] Differential distribution of center-of-mass energies $\sqrt{s_{\overline B B}}$ in $\overline B + B$ annihilations produced per min. bias p+Au reaction.}
    \label{fig:ppbar_sqrts}
\end{figure}
\begin{table}[h!]
    \centering 
    \renewcommand{\arraystretch}{1.5} 
    \begin{tabular}{|l|c|}
         \hline
       Reaction  & Threshold [GeV] \\
         \hline
         \hline
    $ p \overline{p} \rightarrow \Lambda_c^+ \overline{\Lambda_c^-}$ & 4.572  \\ \hline
    $ p \overline{p} \rightarrow \Sigma_c^+ \overline{\Lambda_c^-}$ & 4.804  \\ \hline
    $ p \overline{p} \rightarrow \Sigma_c^+ \overline{\Sigma_c^-}$ & 5.036  \\ \hline
    $ p \overline{p} \rightarrow \Sigma_c^{++} \overline{\Sigma_c^{--}}$ & 5.036 \\ \hline
    $ p \overline{p} \rightarrow \Sigma_c^0 \overline{\Sigma_c^0}$ & 5.036  \\ \hline
    $ p \overline{p} \rightarrow \Xi_c^+ \overline{\Xi_c^-}$ & 4.934  \\ \hline
    $ p \overline{p} \rightarrow \Xi_c^0 \overline{\Xi_c^0}$ & 4.940  \\ \hline
    $ p \overline{p} \rightarrow D^0 \overline{D^0}$ & 3.728  \\ \hline
    $ p \overline{p} \rightarrow D^+ {D^-}$ & 3.738  \\ \hline
    $ p \overline{p} \rightarrow D_s^+ {D_s^-}$ & 3.936  \\ \hline\hline
    $ p \overline{p} \rightarrow \phi  \phi $ & 2.038  \\ \hline
    \end{tabular}
    \caption{Threshold energies for the different charm and strangeness channels in proton-anti-proton annihilations.}
    \label{tab:tab1}
\end{table}
\section{Simulations of the collision spectrum of $\overline B + B$ annihilations}
We employ the Ultra-relativistic Quantum Molecular Dynamics (UrQMD) model \cite{Bleicher:1999xi,Bleicher:2022kcu} (v. 3.4) to calculate minimum bias \footnote{We define min. bias proton+Au reactions in the model as those with an impact parameter of $b\leq6$ fm.} $p+Au$ reactions. The simulations are performed in cascade mode. UrQMD includes the production of anti-baryons via diquark-pair creation in the strings created in the individual proton+proton reactions. The production of anti-baryons in the planned CBM energy region has been investigated in the past (for the reaction $p+p \rightarrow \overline{p} +X$, see \cite{Bleicher:2000gj}, for the production in nuclear systems, see \cite{Sombun:2018yqh}) and compares well to the experimental data. The anti-proton annihilation cross section is fitted to available experimental data. We observe that approx. 25\% of the produced anti-protons are annihilated in the target nucleus in minimum bias p+Au reactions.

\begin{table}[t!]
    \centering 
    \renewcommand{\arraystretch}{1.5} 
    \begin{tabular}{|l|c|c|c|}
         \hline
       Reaction  & \multicolumn{3}{c|}{$\sigma^{p \overline{p}\rightarrow x+y}$ [$\mu b$] at $\varepsilon=25$~MeV} \\
         \hline       
         \hline
          & M/B exch. & Quark model & Ref.\\
         \hline
    $ p \overline{p} \rightarrow \Lambda_c^+ \overline{\Lambda_c^-}$    & 3.33 & 0.97  & \cite{Haidenbauer:2016pva}\\ \hline
    $ p \overline{p} \rightarrow \Sigma_c^+ \overline{\Lambda_c^-}$     & 0.41 & 0.015 & \cite{Haidenbauer:2016pva} \\ \hline
    $ p \overline{p} \rightarrow \Sigma_c^+ \overline{\Sigma_c^-}$      & 0.24 & 0.001 & \cite{Haidenbauer:2016pva}  \\ \hline
    $ p \overline{p} \rightarrow \Sigma_c^{++} \overline{\Sigma_c^{--}}$& 0.86 & 0.001 & \cite{Haidenbauer:2016pva}  \\ \hline
    $ p \overline{p} \rightarrow \Sigma_c^0 \overline{\Sigma_c^0}$      & 0.33 & 0.001 & \cite{Haidenbauer:2016pva}  \\ \hline
    $ p \overline{p} \rightarrow \Xi_c^+ \overline{\Xi_c^-}$            & 0.51 & 0.004 & \cite{Haidenbauer:2016pva}  \\ \hline
    $ p \overline{p} \rightarrow \Xi_c^0 \overline{\Xi_c^0}$            & 0.22 & 0.004 & \cite{Haidenbauer:2016pva}  \\ \hline
    $ p \overline{p} \rightarrow D^0 \overline{D^0}$                    & 0.025 & 0.009&\cite{Haidenbauer:2014rva}  \\ \hline
    $ p \overline{p} \rightarrow D^+ {D^-}$                             & 0.03 & 0.007 &\cite{Haidenbauer:2014rva} \\ \hline
    $ p \overline{p} \rightarrow D_s^+ {D_s^-}$                         & 0.02 & 0.025 &\cite{Haidenbauer:2014rva} \\ \hline\hline
    $ p \overline{p} \rightarrow \phi  \phi$                            & 3 & -- &\cite{JETSET:1998akg} \\ \hline    
    \end{tabular}
    \caption{Charmed baryon production cross sections, taken from \cite{Haidenbauer:2016pva}, and charmed meson production cross sections, taken from \cite{Haidenbauer:2014rva}, both at $\varepsilon=25$~MeV above threshold. The double $\phi$ production cross section is taken from the JETSET data \cite{JETSET:1998akg}.}
    \label{tab:tab2}
\end{table}

Let us start with the analysis of the spectrum of $\bar{B} +B$ annihilations in these proton+Au reactions at three different beam energies. The annihilations stem from anti-baryons that where initially produced and then annihilate within the same target nucleus.  Fig. \ref{fig:ppbar_sqrts} shows the distribution of the center-of-mass energy of all individual anti-baryon+baryon annihilations $\sqrt{s_{\Bar{B}B}}$ per $p+Au$ event. One observes that the majority of annihilations is at energies below 3 GeV, however the spectrum stretches (especially in the case of 30 GeV initial proton energy) up to center of mass energies of $5-6$ GeV. Comparing these energies to the threshold energies for the production of exotica, shown in Table \ref{tab:tab1}, one can understand that these annihilations allow for the production of the $\Lambda_c$ baryon and $D\overline{D}$ as well as more exotic final states.

\section{Production rates for exotic hadron production}
To translate the simulated annihilation spectrum of anti-baryon+baryon annihilations, $dN/d\sqrt{s_{\Bar{B}B}}$, into an estimate for the production cross section of charmed hadrons, we will use the cross sections for charm hadron production in $\overline p + p$.
These cross sections have been explored intensively by Haidenbauer and Krein \cite{Haidenbauer:2014rva,Haidenbauer:2016pva} and are shown in Table \ref{tab:tab2}.
As can be seen, the expected cross sections for charmed hadron production in annihilation differ drastically, depending on whether one employs the M/B exchange or quark model.
We will use both scenarios to obtain the corresponding yields of charmed states from annihilations in the p+Au reactions. 
The cross section for $\phi$ production is taken from the JETSET data \cite{JETSET:1998akg}. 
The charmed (hidden strange) hadron production cross sections can be transformed into a yield per annihilation via $N(x)=\sigma(\overline p p \rightarrow x)/\sigma(\overline p p)^\mathrm{ann}$, assuming that $\sigma(\overline p p)^\mathrm{ann} \approx 50$ mb, in the energy range considered.
This yield per annihilation depends only weakly on the invariant mass, when starting about 25 MeV above the threshold. We assume that the charm (strangeness) production cross sections in anti-baryon+baryon annihilations are the same as in anti-proton+proton annihilations at the same center-of-mass energy.

\begin{table}[t!]
    \centering
    \renewcommand{\arraystretch}{1.5}
    \begin{tabular}{|l|c|c|c|c|}
         \hline
       Hadron  & \multicolumn{4}{c|}{Rate per 90 days p+Au} \\
         \hline
         \hline
         & \multicolumn{2}{c|}{E$_\mathrm{lab}=30A$ GeV} & \multicolumn{2}{c|}{E$_\mathrm{lab}=20A$ GeV} \\ \hline
          & M/B exch. & Quark model & M/B exch. & Quark model \\  
         \hline
    $ \Lambda_c^+ \overline{\Lambda_c^-}$    & $3.1\cdot10^4$ & $9.0\cdot10^3$  & $4.9 \cdot 10^3$ & $1.4 \cdot 10^3$ \\ \hline
    $  \Sigma_c^+ \overline{\Lambda_c^-}$    & $2.2\cdot10^3$ & $8.0\cdot10^1$  & $2.7 \cdot 10^2$ & $1.0 \cdot 10^1$ \\ \hline
    $ \Sigma_c^+ \overline{\Sigma_c^-}$      & $5.8\cdot10^2$ & $2.0\cdot10^0$  & $4.0 \cdot 10^1$ & $0.2 \cdot 10^{0}$ \\ \hline
    $  \Sigma_c^{++} \overline{\Sigma_c^{--}}$& $2.1\cdot10^3$& $2.0\cdot10^0$  & $1.4 \cdot 10^2$ & $0.2 \cdot 10^{0}$ \\ \hline
    $  \Sigma_c^0 \overline{\Sigma_c^0}$      & $8.0\cdot10^2$& $2.0\cdot10^0$  & $5.5 \cdot 10^1$ & $0.2 \cdot 10^{0}$ \\ \hline
    $  \Xi_c^+ \overline{\Xi_c^-}$            & $1.5\cdot10^3$& $1.2\cdot10^1$  & $1.5 \cdot 10^2$ & $1.2 \cdot 10^{0}$ \\ \hline
    $  \Xi_c^0 \overline{\Xi_c^0}$            & $6.5\cdot10^2$& $1.2\cdot10^1$  & $6.5 \cdot 10^1$ & $1.2 \cdot 10^{0}$ \\ \hline
    $  D^0 \overline{D^0}$                    & $3.1\cdot10^3$& $1.1\cdot10^3$  & $1.3 \cdot 10^3$ & $4.6 \cdot 10^2$ \\ \hline
    $  D^+ {D^-}$                             & $3.8\cdot10^3$& $8.8\cdot10^2$  & $1.5 \cdot 10^3$ & $3.6 \cdot 10^2$ \\ \hline
    $  D_s^+ {D_s^-}$                         & $1.2\cdot10^3$& $1.6\cdot10^3$  & $4.9 \cdot 10^2$ & $6.0 \cdot 10^2$ \\ \hline\hline
    $  \phi  \phi$                            & $3.9\cdot10^7$ & --   & $1.7 \cdot 10^7$ & -- \\ \hline
    \end{tabular}
    \caption{Charmed (hidden strange) hadron yields per 90 days at full luminosity in min. bias p+Au reactions at E$_\mathrm{lab}=30 A$~GeV (left columns) and E$_\mathrm{lab}=20 A$~GeV (right columns). }
    \label{tab:tab3}
\end{table}

With the simulated annihilation energy spectrum and for a typical running campaign of 90 days, we obtain the following estimates: Multiplying the number of $B \overline{B}$ annihilations per $p+Au$ event above the threshold (plus $\varepsilon = 25$ MeV) with the production rate of hadron species $h_i$ per annihilation with the design collision rate of 10 MHz $(R=10^7 /s)$ for $p+Au$  with the number of seconds in 90 days (${\Delta t_\mathrm{90\,d} \approx 7.8 \cdot 10^6 s}$) and obtain the rates in Table \ref{tab:tab3} for E$_\mathrm{lab}=30 A$~GeV (left columns) and E$_\mathrm{lab}=20 A$~GeV (right columns). For E$_\mathrm{lab}=10 A$~GeV, the obtained yields are too low to be accessible by the experiment.
\begin{equation}
    N^\mathrm{90\,days}(x) = R \ \Delta t_\mathrm{90\,d} \
    \frac{\sigma^{\Bar{p}p\rightarrow x}}{\sigma^{\Bar{p}p}_\mathrm{ann.}}
    \int\limits_{\sqrt{s_\mathrm{thr}}+\varepsilon}^{\infty} \hspace{-0.3cm} \mathrm{d}\sqrt{s}  \frac{\mathrm{d}N^{p+Au}_\mathrm{\Bar{B}B}}{\mathrm{d}\sqrt{s_{\Bar{B}B}}}
\end{equation}

Thus, the rates do allow for exploratory studies of a multitude of charmed mesons and baryons. Even very exotic states like the $D_s$ and the $\Xi_c$ are in experimental reach. 

\section{Conclusion}
Predictions for charmed hadron and double $\phi$ production in p+Au reactions in the FAIR energy range, E$_\mathrm{lab}=30 A$~GeV  and E$_\mathrm{lab}=20 A$~GeV,  were presented, assuming secondary charm and $\phi$ production by anti-baryon-baryon annhilations. These annihilations may provide a higher yield of charm states and double $\phi$ states than p+p reaction due to the gluon-rich environment created in annihilation events. Our simulations predict a small but still measurable amount of exotic hadrons (e.g. $D_s^+$ and $\Xi_c^+$), which have never been observed at such low collision energies and which should be explored at FAIR. 

\section*{Acknowledgments}
We thank F. Nerling for fruitful discussion.
T.R. acknowledges support through the Main-Campus-Doctus fellowship provided by the Stiftung Polytechnische Gesellschaft (SPTG) Frankfurt am Main and moreover thanks the Samson AG for their support.
The computational resources for this project were provided by the Center for Scientific Computing of the GU Frankfurt and the Goethe--HLR.


\end{document}